\documentclass[twocolumn,showpacs,preprintnumbers,amsmath,amssymb]{revtex4}
\usepackage{graphicx}% Include figure files
\usepackage{dcolumn}% Align table columns on decimal point
\usepackage{bm}% bold math
\usepackage{amssymb,amsmath}

\begin{document}

%\preprint{APS/123-QED}

\title{%
Topological odd-parity superconductors
%with and without time-reversal invariance
%Manuscript Title:\\with Forced Linebreak
}% Force line breaks with \\

\author{Masatoshi Sato}
\email{msato@issp.u-tokyo.ac.jp}
% \altaffiliation[Also at ]{The Institute for Solid State Physics, The
% University of Tokyo.}%Lines break automatically or can be forced with \\
\affiliation{%
The Institute for Solid State Physics, The University of Tokyo,\\
Kashiwanoha 5-1-5, Kashiwa, Chiba, 277-8581, Japan,
}%

%\affiliation{

\date{\today}% It is always \today, today,

\begin{abstract}
In this letter, we investigate topological phases of full-gapped
 odd-parity 
 superconductors, which are
 distinguished by the bulk topological invariants and
 the topologically protected gapless boundary states. 
Using the particle-hole symmetry, we introduce $Z_2$
invariants characterizing topological odd-parity
 superconductors without or with time-reversal invariance.
For odd-parity superconductors, a combination of the inversion and
the $U(1)$ gauge symmetry is manifestly preserved,
and the combined
symmetry enables us to evaluate the $Z_2$ invariants from the
knowledge of the Fermi surface structure. 
Relating the $Z_2$ invariants to other topological invariants, we
 establish characterization of topological odd-parity
 superconductors in terms of the Fermi surface topology.
Simple criteria for topological odd-parity superconductors in various
 dimensions are provided.
Implications of our formulas for nodal odd-parity superconductors are
 also discussed. 
\end{abstract}
\pacs{Valid PACS appear here}% PACS, the Physics and Astronomy
\maketitle

%{\sl Introduction -}
Recently, there has been considerable interest in topological phases which 
are characterized by the bulk topological invariants and the
topologically protected gapless boundary states.
The prototype of the topological phase is the integer quantum Hall states,  
where the band TKNN integers or Chern numbers give the integer quantum
Hall effects \cite{TKNN82}, and they ensure the
stability of the gapless edge states at the same time \cite{Hatsugai93}.  
While the time-reversal symmetry breaking (TRSB) is necessary to have non-trivial Chern
numbers, there exist another topological invariants called the $Z_2$
invariants which classify the 
topological phases of the time-reversal invariant (TRI) insulators
\cite{KM05a,MB07,Roy09,FK07}. 
When the $Z_2$ invariants are non-trivial, there
exist an odd number of the Kramers pairs of gapless edge modes in two
dimensions, 
and an odd number of the Kramers degenerate band crossings (Dirac cones)
on the surface in three dimensions,
respectively. 

The concept of topological phases is also applicable to
superconducting states \cite{RG00,Sato06,QHRZ09,SRFL08,SF09,STF09} because
there is a direct analogy between superconductors and insulators:
The Bogoliubov de-Gennes (BdG) Hamiltonian for a quasiparticle of a
superconductor is analogous to the Hamiltonian of a band insulator,
and the superconducting gap corresponds to the gap of the band insulator.
Indeed, the TRSB chiral $p$-wave superconductors have
non-trivial Chern numbers, and they support topologically protected
chiral gapless edge states in analogy with the integer quantum Hall
states \cite{RG00}.
Topological phases of noncentrosymmetric superconductors and
s-wave superfluids, which support non-abelian anyons, were also
investigated in \cite{Sato06,SF09,STF09}. 

In addition to the analogous properties, there are topological features
inherent to superconductors.
Superconductors possess the particle-hole
symmetry (PHS) exchanging the quasiparticle with the
anti-quasiparticle, which provides additional topological characteristics to
superconductors. In particular, for general superconductors without spin
rotation symmetry,
there arise extra $Z_2$ invariants in one dimensional TRSB and TRI
systems, and an integer winding number in three dimensional TRI one
\cite{SRFL08}.
As a result, the topological superconductors are
characterized by the one dimensional $Z_2$ invariants and the two
dimensional Chern number for the TRSB case, and the one and two dimensional
$Z_2$ invariants and the three dimensional winding number for the TRI one,
respectively.  

In this letter, assuming the inversion symmetry in the normal state,
we present a theory of topological odd-parity superconductors.
For TRI single-band odd-parity superconductors, it has been
revealed that the topological properties are characterized by the Fermi
surface topology in the normal state \cite{Sato09}.
Here we extend these results 
to general odd-parity superconductors without or with time-reversal
invariance,
by using the one-dimensional $Z_2$ invariants obtained from the PHS.
We develop a method to link the $Z_2$ invariants
to the topology of the Fermi surface,
where a combination of the
inversion and the $U(1)$ gauge symmetry
plays an essential role.
Moreover, making connections between the $Z_2$ invariants and the
other topological invariants mentioned above,  
we provide characterization of topological odd-parity
superconductors in terms of the topology of the Fermi surface.

In the following, we consider a general Hamiltonian $H$ \footnote{
In the classification of \cite{SRFL08},
the Hamiltonian is in class D
class for the TRSB case,
and in class DIII for TRI one, respectively.
} for full
gapped odd-parity superconducting states,
\begin{eqnarray}
&&H=\frac{1}{2}\sum_{{\bm k}\alpha\alpha'}
(c_{{\bm k}\alpha}^{\dagger}, c_{-{\bm k}\alpha})
H({\bm k})
\left(
\begin{array}{c}
c_{{\bm k}\alpha'} \\
c^{\dagger}_{-{\bm k}\alpha'}
\end{array}
\right),
\nonumber\\
&&H({\bm k})=\left(
\begin{array}{cc}
{\cal E}({\bm k})_{\alpha\alpha'} &\Delta({\bm k})_{\alpha\alpha'} \\
\Delta^{\dagger}({\bm k})_{\alpha\alpha'} & -{\cal E}^{T}(-{\bm
 k})_{\alpha\alpha'}
\end{array}
\right),
\label{eq:hamiltonian}
\end{eqnarray}
where $c^{\dagger}_{{\bm k}\alpha}$ ($c_{{\bm k}\alpha}$) denotes the
creation (annihilation) operator of electron with momentum ${\bm k}$.
The suffix $\alpha$ labels other degrees of freedom for electron such as
spin, orbital degrees of freedom, sub-lattice indeces, and so on. 
${\cal E}({\bm k})$ is an hermitian matrix describing the normal
dispersion of the electron. Here we assume that the system in the normal
state is symmetric under the inversion  
$
c_{{\bm k}\alpha}\rightarrow \sum_{\alpha'}
P_{\alpha \alpha'}c_{-{\bm k}\alpha'}$ with
$P^2=1$, 
so
$
P^{\dagger} {\cal E}({\bm k}) P={\cal E}(-{\bm k}).
$
For an odd-parity superconductor, the gap function $\Delta({\bm k})$
satisfies 
$
P^{\dagger}\Delta({\bm k})P^{*}=-\Delta(-{\bm k}). 
$
In addition, the Fermi statistics of electron implies 
$
\Delta^{T}({\bm k})=-\Delta(-{\bm k}). 
$

An important ingredient of our theory is the 
PHS of the BdG Hamiltonian (\ref{eq:hamiltonian}), 
\begin{eqnarray}
CH({\bm k})C^{\dagger}=-H^{*}(-{\bm k}), 
\quad
C=\left(
\begin{array}{cc}
0 &1 \\
1 & 0
\end{array}
  \right). 
\label{eq:particle-hole}
\end{eqnarray}
From (\ref{eq:particle-hole}), we can say that if $|u_n({\bm k})\rangle$
is a quasiparticle state with positive energy $E_n({\bm k})>0$ satisfying 
$
H({\bm k})|u_n({\bm k})\rangle=E_n({\bm k})|u_n({\bm k})\rangle, 
$
then $C|u_n^{*}(-{\bm k})\rangle$ is a quasiparticle state with
negative energy $-E_n(-{\bm k})<0$.
In the following, we use a positive (negative) $n$ for  $|u_n({\bm k})\rangle$
to represent
a positive (negative) energy quasiparticle state, and set 
\begin{eqnarray}
|u_{-n}({\bm k})\rangle =C |u_n^{*}(-{\bm k})\rangle. 
\label{eq:negative energy state}
\end{eqnarray}

To define the topological invariants, we introduce the gauge fields
$
A^{(\pm )}_i({\bm k})=i\sum_{n\gtrless 0}
\langle u_n({\bm k})|\partial_{k_i}|u_n({\bm k})\rangle 
$.
We also denote their sum as $A_i({\bm k})=A_i^{(+)}({\bm
k})+A_i^{(-)}({\bm k})$. 
From (\ref{eq:particle-hole}), we have
\begin{eqnarray}
A^{(+)}_i({\bm k})=A^{(-)}_i(-{\bm k}). 
\label{eq:a+a-}
\end{eqnarray}
Using this and the fact that $A_i({\bm k})$ is the total derivative of a
function, 
we can prove that the Wilson loop of $A_i^{(-)}({\bm k})$ along the
TRI closed path ${\rm C}$ in the Brillouin zone (BZ)
\begin{eqnarray}
W[{\rm C}]=\frac{1}{2\pi}\oint_{\rm C}dk_i A_i^{(-)}({\bm k})
\label{eq:wilsonloop}
\end{eqnarray}
is quantized as 
$
e^{2\pi iW[{\rm C}] }=\pm 1
$ \cite{QHZ08}.
Therefore, we can introduce a $Z_2$ invariant $(-1)^{\nu[{\rm
C}]}$ by $(-1)^{\nu[{\rm C}]}\equiv e^{2\pi iW[{\rm C}]}$.
As we will discuss later, 
$(-1)^{\nu[{\rm C}]}=-1$ (+1)  
corresponds to a topological non-trivial (trivial) phase of
the superconducting state.

\begin{figure}[h]
\begin{center}
\includegraphics[width=6.5cm]{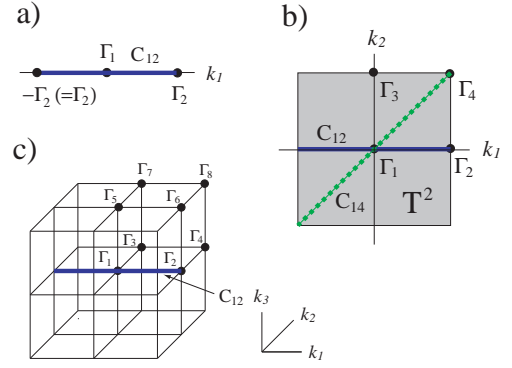}
\caption{The TRI momenta $\Gamma_i$, and the TRI closed path ${\rm
 C}_{ij}$ connecting $\Gamma_i$ and $\Gamma_j$ in the BZ. a) 1 dim BZ.
The solid line denotes ${\rm C}_{12}$.
b) 2 dim BZ $T^2$. The solid line denotes ${\rm C}_{12}$, and the
 dotted one ${\rm C}_{14}$. c) 3 dim BZ. The solid line is ${\rm C}_{12}$.}
\label{fig:cij}
\end{center}
\end{figure}

Now consider the TRI closed path ${\rm C}_{ij}$
passing through the TRI momenta $\Gamma_i$ and $\Gamma_j$ in Fig.\ref{fig:cij}.
The TRI momentum satisfies $\Gamma_i=-\Gamma_i+{\bm G}$ with a
reciprocal lattice vector ${\bm G}$, and because of the periodicity of
the BZ, ${\rm C}_{ij}$ forms a closed path.
In the following, we evaluate
the $Z_2$ invariant
$(-1)^{\nu[{\rm C}_{ij}]}$ along ${\rm C}_{ij}$ by developing the
method in
\cite{Sato09}.

For an odd-parity superconductor,
the combination of the inversion
symmetry and the $U(1)$ gauge symmetry,  
$
c_{{\bm k}\alpha}\rightarrow iP_{\alpha\alpha'}c_{{\bm k}\alpha'},
$
is manifestly preserved, 
although each
symmetry is spontaneously broken by the
condensation $\Delta({\bm k})$.
Therefore, the BdG Hamiltonian $H({\bm k})$
has the following symmetry
\begin{eqnarray}
\Pi^{\dagger} H({\bm k}) \Pi=H(-{\bm k}),
\quad
\Pi=
\left(
\begin{array}{cc}
P & 0\\
0 & -P^{*}
\end{array}
\right). 
\end{eqnarray}
From this symmetry, we have $[H({\Gamma_i}), \Pi]=0$ for the TRI
momentum $\Gamma_i$. Thus, the quasiparticle state
$|u_n(\Gamma_i)\rangle$ at $\Gamma_i$ is simultaneously an eigenstate of $\Pi$,
$
\Pi |u_n(\Gamma_i)\rangle=\pi_n(\Gamma_i)|u_n(\Gamma_i)\rangle. 
$
Evaluation of $(-1)^{\nu[{\rm C}_{ij}]}$ is done by using the unitary
matrices,
$
V_{mn}({\bm k})=\langle u_m({\bm k}) |\Pi C |u_n^{*}({\bm k})\rangle,
$
and
$
W_{mn}({\bm k})=\langle u_m(-{\bm k})|C|u_n^{*}({\bm k})\rangle. 
$
Since we have
${\rm tr}(V^{\dagger}\partial_{k_i}V)=2iA_i({\bm k})$
from (\ref{eq:negative energy state}),
$\nu[{\rm C}_{ij}]$ is rewritten as
\begin{eqnarray}
\nu[{\rm C}_{ij}]
=\frac{1}{\pi}\int_{\Gamma_i}^{\Gamma_j}dk_i A_i({\bm k}) 
=\frac{1}{\pi i} \ln \left(
 \frac{\sqrt{{\rm det}V(\Gamma_i)}}{\sqrt{{\rm det}V(\Gamma_j)}}
\right).
\label{eq:Vformula}
\end{eqnarray}
Here we have used (\ref{eq:a+a-}) and 
$
{\rm tr}(V^{\dagger}\partial_{k_i}V)=\partial_{k_i}\ln {\rm det}V.
$
Furthermore, because $V_{mn}(\Gamma_i)$ is recast into
$
V_{mn}(\Gamma_i)=\langle u_m(\Gamma_i)|\Pi C|u_n^{*}(\Gamma_i)\rangle
=\pi_m(\Gamma_i)\langle u_m(\Gamma_i)|C|u_n^{*}(\Gamma_i)\rangle
=\pi_m(\Gamma_i)W_{mn}(\Gamma_i), 
$
we obtain 
\begin{eqnarray}
{\rm det}V(\Gamma_i)=\prod_n \pi_n(\Gamma_i){\rm det}W, 
\label{eq:prod}
\end{eqnarray}
where ${\rm det}W$ is
independent of $\Gamma_i$ 
because
$
\partial_{k_i}\ln {\rm det}W=
{\rm tr}(W^{\dagger}\partial_{k_i} W)
=i[A_i({\bm k})-A_i(-{\bm k})]=0.
$
Due to the PHS, $|u_n(\Gamma_i)\rangle$ and
$|u_{-n}(\Gamma_i)\rangle$ share the same eigenvalue of $\Pi$ 
and each eigenvalue appears twice in the product in (\ref{eq:prod}).
Therefore, taking the square root, we find
$
\sqrt{{\rm det}V(\Gamma_i)}=\prod_{n<0}\pi_n(\Gamma_i)\sqrt{{\rm det}W}. 
$
As a result, (\ref{eq:Vformula}) reduces to
\begin{eqnarray}
(-1)^{\nu[{\rm C}_{ij}]}=\prod_{n<0}\pi_n(\Gamma_i)\pi_n(\Gamma_j),
\label{eq:piformula}
\end{eqnarray}
where we have used $\pi_n^2(\Gamma_j)=1$.

In order to attribute 
the Fermi surface properties to the $Z_2$ invariants,
we make the weak-paring assumption \cite{QHZ09}. 
For ordinary superconductors, the superconducting gap is much smaller than the
Fermi energy.
Therefore, we reasonably assume that  the typical energy scale of
the gap function $\Delta(\Gamma_i)$ at the TRI
momentum is much smaller than that of ${\cal E}(\Gamma_i)$.  
Under this assumption, we can take $\Delta({\Gamma_i})\rightarrow 0$
without the gap closing.
Because of the topological nature of $(-1)^{\nu[{\rm C}_{ij}]}$, this
adiabatic process does not change the value of $(-1)^{\nu[{\rm C}_{ij}]}$.

In the process $\Delta(\Gamma_i)\rightarrow 0$, the BdG Hamiltonian at
$\Gamma_i$ reduces to
$H(\Gamma_i)\rightarrow {\rm diag}({\cal E}(\Gamma_i),-{\cal E}^{T}(\Gamma_i))$.
By using an eigenstate $|\varphi(\Gamma_i)\rangle$ of ${\cal
E}(\Gamma_i)$ satisfying
$
{\cal E}(\Gamma_i)|\varphi_{\alpha}(\Gamma_i)\rangle
=\varepsilon_{\alpha}(\Gamma_i)|\varphi_{\alpha}(\Gamma_i)\rangle, 
$
an occupied state of $H(\Gamma_i)$ is given by
$(|\varphi_{\alpha}(\Gamma_i)\rangle,0)^{t}$ for
$\varepsilon_{\alpha}(\Gamma_i)<0$, and
$(0, |\varphi^*_{\alpha}(\Gamma_i)\rangle)^{t}$ for
$\varepsilon_{\alpha}(\Gamma_i)>0$.
Therefore, denoting the parity of $|\varphi_{\alpha}(\Gamma_i)\rangle$ as
$
P|\varphi_{\alpha}(\Gamma_i)\rangle
=\xi_{\alpha}(\Gamma_i)|\varphi_{\alpha}(\Gamma_i)\rangle,  
$
we find 
\begin{eqnarray}
\prod_{n<0}\pi_n(\Gamma_i)=\prod_{\alpha}\xi_{\alpha}(\Gamma_i)\prod_{\alpha}
 {\rm sgn}\varepsilon_{\alpha}(\Gamma_i),
\label{eq:piepsilon}
\end{eqnarray}
where the sum of $\alpha$ is taken for all
eigenstates of ${\cal E}({\Gamma_i})$.
We notice here that the product of the parity,
$\prod_{\alpha}\xi_{\alpha}(\Gamma_i)$, is independent of $\Gamma_i$
since it is determined solely from ${\rm det}P$ and the dimensionality of
the matrix ${\cal E}({\bm k})$. 
Thus if we substitute (\ref{eq:piepsilon}) into (\ref{eq:piformula}),
the contributions from the parity at $\Gamma_i$ and $\Gamma_j$ cancel
each other, then we obtain the final expression,  
\begin{eqnarray}
(-1)^{\nu[{\rm C}_{ij}]}=\prod_{\alpha}{\rm
sgn}\varepsilon_{\alpha}(\Gamma_i)
{\rm sgn}\varepsilon_{\alpha}(\Gamma_j).
\label{eq:epsilonformula}
\end{eqnarray}
This formula is very powerful: It enables us to calculate the $Z_2$
invariants only from
the knowledge of the band energy $\varepsilon_{\alpha}(\Gamma_i)$ of
electron in the normal state.
We also notice that the right hand side of (\ref{eq:epsilonformula}) has
its own topological meaning.
By denoting the number of intersection points between the Fermi surface
and ${\rm C}_{ij}$ as $i_0(S_{\rm F}\cap {\rm C}_{ij})$, the right hand side is
found to be $(-1)^{i_0(S_{\rm F}\cap {\rm C}_{ij})/2}$ \footnote{In this
letter, the
degeneracy of the Fermi surface is taken into account to count 
$i_0(S_{\rm F}\cap{\rm C}_{ij})$, $p_0(S_{\rm F})$ and $\chi(S_{\rm
F})$.
So, they are different from those in \cite{Sato09} by factor 2.}.
Therefore, when $i_0(S_{\rm F}\cap {\rm C}_{ij})$ is odd (even), the
$Z_2$ invariant
in (\ref{eq:epsilonformula}) is non-trivial (trivial).

\begin{figure}[t]
\begin{center}
\includegraphics[width=5cm]{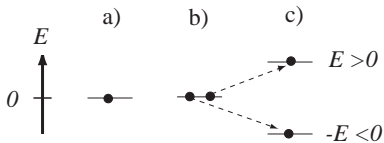}
\caption{$Z_2$ classification of edge state. a) Topologically
 protected zero mode. b) and c) Topologically trivial edge modes.}
\label{fig:particlehole}
\end{center}
\end{figure}

To see the physical meaning of the $Z_2$ invariants, consider
a full gapped one-dimensional odd-parity superconductor.
In one dimension, we have a single
$Z_2$ invariant $(-1)^{\nu[{\rm C}_{12}]}$ where $C_{12}$ is the TRI
closed path
in Fig.\ref{fig:cij}a).
When 
$(-1)^{\nu[{\rm C}_{12}]}= -1$ (+1),
the system is topologically non-trivial (trivial), and 
there exist an odd (even) number of the zero energy
states on the boundary \cite{QHZ08}.
Without loss of generality, we can consider the simplest non-trivial
topological phase with a single boundary zero mode.
See Fig.\ref{fig:particlehole}a).
The PHS ensures the topological stability of the zero
mode against small perturbation:
From the PHS,
a non-zero mode with energy $E$ must be paired with a non-zero mode having
the opposite energy $-E$.
Therefore,
a zero mode can be
a non-zero mode only in pairs.
This implies that the single zero mode in Fig.\ref{fig:particlehole}a) can
not aquire non-zero energy and it is stable against small perturbation.
The $Z_2$ nature of the topological phase is evident if we add another
zero mode as depicted in
Fig.\ref{fig:cij}b). In this case, we have a pair of zero modes,
so they can be non-zero modes by small perturbation as demonstrated in
Fig.\ref{fig:cij}c).
This argument consistently indicates that the topological phase
ensured by the PHS is distinguished by the $Z_2$ invariant.

In two and three dimensions, we have multiple one-dimensional $Z_2$
invariants corresponding to possible ${\rm C}_{ij}$ in
Fig.\ref{fig:cij}b)-c).
When the $Z_2$ invariant $(-1)^{\nu[{\rm C}_{ij}]}$ is non-trivial,
we have a gapless state on the surface perpendicular to ${\rm C}_{ij}$: 
By fixing the momenta along the surface perpendicular to ${\rm C}_{ij}$,
the part of the system can be considered as a one-dimensional gapful system
\cite{RH02}.
Therefore, from the topological argument above, it is concluded that
there exist a
gapless mode on the surface.

Following the classification in \cite{SRFL08}, 
the two-dimensional Chern numbers $\nu_{\rm Ch}$ also characterize the
topological phase of the TRSB superconductors. 
Now make a connection between the Chern number $\nu_{\rm Ch}$
and the $Z_2$ invariants in two dimensions.
$\nu_{\rm Ch}$ is defined by 
\begin{eqnarray}
\nu_{\rm Ch}=\frac{1}{2\pi}\int_{T^2}{\cal F}^{(-)}({\bm k}), 
\end{eqnarray}
where ${\cal F}^{(-)}({\bm k})$ is the field strength of $A_i^{(-)}({\bm
k})$, and $T^2$ the two-dimensional BZ in Fig.\ref{fig:cij}b) \cite{TKNN82}.
Noting that the field strength ${\cal F}({\bm k})$ of $A_i({\bm k})$
is identically zero,   
we find
$
{\cal F}^{(-)}({\bm k})={\cal F}^{(-)}(-{\bm k}) 
$
from (\ref{eq:a+a-}).
Thus the Chern number is linked to the $Z_2$ invariants as
\begin{eqnarray}
\nu_{\rm Ch}&=&\frac{1}{2\pi}\int_{T^2}{\cal F}^{(-)}({\bm k}) 
= \frac{1}{\pi}\int_{T^2_{+}}{\cal F}^{(-)}({\bm k}) 
\nonumber\\
&=& \frac{1}{\pi}\oint_{\partial T_{+}^2}dk_i A_i^{(-)}({\bm k})
=\nu[{\rm C}_{12}]-\nu[{\rm C}_{34}],
\label{eq:chw}
\end{eqnarray}
where $T_{+}^2$ is the upper half of $T^2$. Consequently, 
from (\ref{eq:epsilonformula}), we have the following relation between
$\nu_{\rm CH}$ and the topology of the Fermi surface,
\begin{eqnarray}
(-1)^{\nu_{\rm Ch}}
=\prod_{\alpha, i=1,2,3,4}{\rm
sgn}\varepsilon_{\alpha}(\Gamma_i)=(-1)^{p_0(S_{\rm F})}, 
\label{eq:chformula}
\end{eqnarray}
where $p_0(S_{\rm F})$ is the number of the connected components of the Fermi
surface on $T^2$, and 
in the second equality, we have used the result in \cite{Sato09}. 
This formula provide a criterion for non-zero $\nu_{\rm Ch}$:
If $p_0(S_{\rm F})$ is odd, then $\nu_{\rm Ch}$ is non-zero.
This simple criterion immediately reproduces the non-zero $\nu_{\rm
Ch}$ for the chiral $p$-wave superconductor \cite{RG00} since it has a
single Fermi surface. 
In a similar manner, 
it is found that the Chern numbers in three dimensions
are also characterized by the topology of the Fermi surface.

Now consider TRI odd-parity superconductors.
Because of the time-reversal invariance $\Theta$ with $\Theta^2=-1$,
the occupied states of the BdG
Hamiltonian $H({\bm k})$ are divided into Kramers pairs,
$|u_n^{s}({\bm k})\rangle$ $(s={\rm I},{\rm II})$,
\begin{eqnarray}
|u_n^{\rm I}({\bm k})\rangle =\Theta |u_n^{{\rm II}}(-{\bm k})\rangle. 
\label{eq:Kramers}
\end{eqnarray}
Since the Kramers pair, $|u^{{\rm I}}_{n}(\Gamma_i)\rangle \equiv
|u_{2n}(\Gamma_i)\rangle$ and
$|u^{{\rm II}}_{n}(\Gamma_i)\rangle\equiv |u_{2n+1}({\Gamma_i})\rangle$
at $\Gamma_i$ share the same
eigenvalue of $\Pi$,  (\ref{eq:piformula}) leads to $(-1)^{\nu[{\rm
C}_{ij}]}=1$.
In other words, $(-1)^{\nu[{\rm C}_{ij}]}$ is always trivial for
TRI odd-parity superconductors.
However, use of the time-reversal invariance as well makes it
possible to define non-trivial $Z_2$ invariants.

To define non-trivial $Z_2$ invariants, let us introduce
the gauge field $A_i^{s(-)}({\bm k})=i\sum_{n<0}\langle u_n^{s}({\bm k})|
\partial_{k_i}|u_n^{s}({\bm k})\rangle$, 
and  its Wilson loop $W^{s}[{\rm C}_{ij}]$ for the Kramers pairs $|u_n^{s}({\bm
k})\rangle$,
\begin{eqnarray}
W^{s}[{\rm C}_{ij}]=\frac{1}{2\pi}\oint_{{\rm C}_{ij}} dk_i
 A_i^{s(-)}({\bm k}).
\label{eq:halfwilson}
\end{eqnarray}
Because
$W[{\rm C}_{ij}]$ is divided into 
$
W[{\rm C}_{ij}]=W^{{\rm I}}[{\rm C}_{ij}]+W^{{\rm II}}[{\rm C}_{ij}],  
$
and
$W^{{\rm I}}[{\rm C}_{ij}]=W^{\rm II}[{\rm C}_{ij}] $ from
(\ref{eq:Kramers}), 
we find
$
W^{\rm I}[{\rm C}_{ij}]
=\nu[{\rm C}_{ij}]/4.
$
Therefore, using $(-1)^{\nu[{\rm C}_{ij}]}=1$, we obtain the
quantization of $e^{2\pi i
W^{\rm I}[{\rm C}_{ij}]}$ as $e^{2\pi i
W^{\rm I}[{\rm C}_{ij}]}=\pm 1$.
This means that
different $Z_2$
invariants $(-1)^{\tilde{\nu}[{\rm C}_{ij}]}$ can be introduced by 
$(-1)^{\tilde{\nu}[{\rm C}_{ij}]}=e^{2\pi i W^{\rm I}[{\rm C}_{ij}]}$.  

For these $Z_2$ invariants $(-1)^{\tilde{\nu}[{\rm C}_{ij}]}$,
it is shown that
\begin{eqnarray}
(-1)^{\tilde{\nu}[{\rm C}_{ij}]}
=
\prod_{\alpha}{\rm sgn}\varepsilon_{2\alpha}(\Gamma_i)
{\rm sgn}\varepsilon_{2\alpha}(\Gamma_j),
\label{eq:epsilonformula2}
\end{eqnarray}
under the same weak-paring assumption as (\ref{eq:epsilonformula}).
Here  $\varepsilon_{\alpha}(\Gamma_i)$ is an eigenvalue of ${\cal
E}(\Gamma_i)$, and we have set
$\varepsilon_{2\alpha}(\Gamma_i)=\varepsilon_{2\alpha+1}(\Gamma_i)$ by
using the Kramers degeneracy. 
In terms of the topology of the Fermi surface, 
the right hand side of (\ref{eq:epsilonformula2}) is expressed by
$(-1)^{i_0(S_{\rm F}\cap {\rm C}_{ij})/4}$.

For the TRI odd-parity superconductors, we also have two-dimensional $Z_2$
invariants $(-1)^{\nu_{\rm 2dTI}}$ which were originally used to characterize
topological insulators \cite{KM05a}, and three dimensional winding
number $\nu_{\rm w}$ defined in \cite{SRFL08}. 
Using the formula (\ref{eq:epsilonformula2}), we can connect these
topological invariants to the Fermi surface topology:
First, since the $Z_2$ invariant $(-1)^{\nu_{\rm 2dTI}}$ for topological
insulators is defined
as a product of $e^{2\pi iW^{\rm I}[{\rm C}_{ij}]}$ \cite{FK07},
it is also a product of our $Z_2$ invariants
$(-1)^{\nu[{\rm C}_{ij}]}$.
Therefore, in two dimensions, the formula  (\ref{eq:epsilonformula2}) leads to 
\begin{eqnarray}
(-1)^{\nu_{\rm 2dTI}}=\prod_{\alpha,i=1,2,3,4}{\rm sgn}
\varepsilon_{2\alpha}(\Gamma_i)=(-1)^{p_0(S_{\rm F})/2}.
\label{eq:2dTI}
\end{eqnarray}
In a similar manner, the $Z_2$ invariant $(-1)^{\nu_{\rm
3dTI}}$ for three dimensional topological insulators
\cite{MB07,Roy09,FK07} is represented by a product of 
our one-dimensional $Z_2$ invariants $(-1)^{\nu[{\rm C}_{ij}]}$.
Moreover, the winding number $\nu_{\rm w} $
satisfies $(-1)^{\nu_{\rm w}}=(-1)^{\nu_{\rm 3dTI}}$
\cite{Sato09}. Accordingly, we have
\begin{eqnarray}
(-1)^{\nu_{\rm w}}= \prod_{\alpha, i=1,\cdots,8}{\rm
 sgn}\varepsilon_{2\alpha}(\Gamma_i)=(-1)^{\chi(S_{\rm F})/4},
\label{eq:winding}
\end{eqnarray}
where $\chi(S_{\rm F})$ is the Euler characteristics of the Fermi
surface \cite{Sato09}.
While the formulas (\ref{eq:2dTI}) and (\ref{eq:winding}) have already
been reported for
TRI single-band spin-triplet superconductors and multi-band
odd-parity
superconductors with $P=1$ \cite{Sato09}, here they are extended to
general TRI odd-parity superconductors
\footnote{The recent preprint \cite{FB09} also discussed the
generalization of (\ref{eq:winding}) for the TRI case in a different
manner.}.
Furthermore, owing to the PHS, we have an additional formula (\ref{eq:epsilonformula2}), which was not known before.

So far, we have considered full-gapped odd-parity superconductors, but 
our formulas (\ref{eq:epsilonformula}) and (\ref{eq:epsilonformula2})
are applicable to a nodal odd parity superconductor as well if the TRI path ${\rm
C}_{ij}$ does not intersect a node of the
superconducting gap.
As was discussed above, fixing the momenta along the surface
perpendicular to ${\rm C}_{ij}$, we can consider 
the part of the system as a one-dimensional gapful odd-parity superconductor. 
When the $Z_2$ invariant $(-1)^{\nu[{\rm C}_{ij}]}$ or
$(-1)^{\tilde{\nu}[{\rm C}_{ij}]}$ is non-trivial, a
gapless surface state is predicted on the surface perpendicular to ${\rm
C}_{ij}$.  

To conclude, in this letter, we present a description of topological
odd-parity superconductors in terms of the Fermi surface topology in the
normal state.
All the topological invariants for
odd-parity superconductors are directly related to the topology of
the Fermi surface by
(\ref{eq:epsilonformula}) and (\ref{eq:chformula}) for the TRSB case,
and (\ref{eq:epsilonformula2}), (\ref{eq:2dTI}) and (\ref{eq:winding})
for the TRI one, respectively, which provide simple criteria for
topological odd-parity superconductors.

The author thanks 
Y.~Asano, S.~Fujimoto, M.~Oshikawa, T.~Takimoto, 
Y.~Tanaka, and X.~Wan for helpful discussions.
I am also grateful for the hospitality for APCTP, Korea, where a
part of this work was done.
This work was partly supported by the Sumitomo Foundation.

\bibliography{topological_order}% Produces the bibliography via BibTeX.

\end{document}